\documentclass[superscriptaddress,prd,showpacs,twocolumn,showpacs,nofootinbib]{revtex4-1}
\usepackage{graphicx}
\usepackage{color}
\usepackage{epsf}
\usepackage{bm}
\usepackage{amsmath}
\usepackage{amsfonts}
\usepackage{amssymb}
\usepackage{epstopdf}%
\usepackage{hyperref}

\newcommand{\lp}{\left(}
\newcommand{\rp}{\right)}
\newcommand{\lb}{\left[}
\newcommand{\rb}{\right]}

\newcommand{\lsim}   {\mathrel{\mathop{\kern 0pt \rlap
  {\raise.2ex\hbox{$<$}}}
  \lower.9ex\hbox{\kern-.190em $\sim$}}}
\newcommand{\gsim}   {\mathrel{\mathop{\kern 0pt \rlap
  {\raise.2ex\hbox{$>$}}}
  \lower.9ex\hbox{\kern-.190em $\sim$}}}
\newcommand{\bw}{\begin{widetext}\begin{equation}}
\newcommand{\ew}{\end{equation}\end{widetext}}
\newcommand{\be}{\begin{equation}}
\newcommand{\ee}{\end{equation}}
\newcommand{\ba}{\begin{eqnarray}}
\newcommand{\ea}{\end{eqnarray}}

\newcommand{\nn}{\nonumber}

\begin{document}

%%%%%%%%%%%%%%%%%%%%%%%%%%%%%%%%%%%%%%%%%%%%%%%%%%%%%%%%%%%%%%%%%%%%%%%%%%%%%%%%%%%%
\title{Coupling matter in modified $Q$-gravity}
%%%%%%%%%%%%%%%%%%%%%%%%%%%%%%%%%%%%%%%%%%%%%%%%%%%%%%%%%%%%%%%%%%%%%%%%%%%%%%%%%%%%

%%%%%%%%%%%%%%%%%%%%%%%%%%%%%%%%%%%%%%%%%%%%%%%%%%%%%%%%%%%%%%%%%%%%%%%%%%%%%%%%%%%%
\author{Tiberiu Harko}
\email{t.harko@ucl.ac.uk}
\affiliation{Department of Physics, Babes-Bolyai University, Kogalniceanu Street,
Cluj-Napoca 400084, Romania,}
\affiliation{School of Physics, Sun Yat-Sen University,   Guangzhou 510275, People's
Republic of China}
\affiliation{Department of Mathematics, University College London, Gower Street, London
WC1E 6BT, United Kingdom}

\author{Tomi S. Koivisto}
\email{timoko@kth.se}
\affiliation{Nordita, KTH Royal Institute of Technology and Stockholm University, Roslagstullsbacken 23, 10691 Stockholm, Sweden}

\author{Francisco S. N. Lobo}
\email{fslobo@fc.ul.pt}
\affiliation{Instituto de Astrof\'{\i}sica e Ci\^{e}ncias do Espa\c{c}o, Faculdade de
Ci\^encias da Universidade de Lisboa, Edif\'{\i}cio C8, Campo Grande,
P-1749-016 Lisbon, Portugal}
\affiliation{Departamento de F\'{\i}sica, Faculdade de Ci\^{e}ncias da Universidade de Lisboa, Edif\'{\i}cio C8, Campo Grande, P-1749-016 Lisboa, Portugal}

\author{Gonzalo J. Olmo}
\email{gonzalo.olmo@uv.es}
\affiliation{Departamento de F\'{i}sica Te\'{o}rica and IFIC, Centro Mixto Universidad de Valencia - CSIC.
Universidad de Valencia, Burjassot-46100, Valencia, Spain}
\affiliation{Departamento de F\'isica, Universidade Federal da
Para\'\i ba, 58051-900 Jo\~ao Pessoa, Para\'\i ba, Brazil}

\author{Diego Rubiera-Garcia}
\email{drgarcia@fc.ul.pt}
\affiliation{Instituto de Astrof\'{\i}sica e Ci\^{e}ncias do Espa\c{c}o, Faculdade de
Ci\^encias da Universidade de Lisboa, Edif\'{\i}cio C8, Campo Grande,
P-1749-016 Lisbon, Portugal}
%%%%%%%%%%%%%%%%%%%%%%%%%%%%%%%%%%%%%%%%%%%%%%%%%%%%%%%%%%%%%%%%%%%%%%%%%%%%%%%%%%%%

\date{\today}

%%%%%%%%%%%%%%%%%%%%%%%%%%%%%%%%%%%%%%%%%%%%%%%%%%%%%%%%%%%%%%%%%%%%%%%%%%%%%%%%%%%%
\begin{abstract}

We present a novel theory of gravity by considering an extension of symmetric teleparallel gravity. This is done by introducing, in the framework of the metric-affine formalism, a new class of theories where the nonmetricity $Q$ is nonminimally coupled to the matter Lagrangian.
More specifically, we consider a Lagrangian of the form $L \sim f_1(Q) + f_2(Q) L_M$, where $f_1$ and $f_2$ are generic functions of $Q$, and $L_M$ is the matter Lagrangian. This nonminimal coupling entails the nonconservation of the energy-momentum tensor, and consequently the appearance of an extra force. The motivation is to verify whether the subtle improvement of the geometrical formulation, when implemented in the matter sector, would allow more universally consistent and viable realisations of the nonminimal curvature-matter coupling theories. Furthermore, we consider several cosmological applications by presenting the evolution equations and imposing specific functional forms of the functions $f_1(Q)$ and $f_2(Q)$, such as power-law and exponential dependencies of the nonminimal couplings.
Cosmological solutions are considered in two general classes of models, and found to feature accelerating expansion at late times.
%A plethora of cosmological scenarios are obtained, and in all considered models, the universe attains an exponentially accelerating de Sitter phase in the late time limit.

\end{abstract}
%%%%%%%%%%%%%%%%%%%%%%%%%%%%%%%%%%%%%%%%%%%%%%%%%%%%%%%%%%%%%%%%%%%%%%%%%%%%%%%%%%%%

%%%%%%%%%%%%%%%%%%%%%%%%%%%%%%%%%%%%%%%%%%%%%%%%%%%%%%%%%%%%%%%%%%%%%%%%%%%%%%%%%%%%
%%%%%%%%%%%%%%%%%%%%%%%%%%%%%%%%%%%%%%%%%%%%%%%%%%%%%%%%%%%%%%%%%%%%%%%%%%%%%%%%%%%%

\maketitle
%\tableofcontents

%%%%%%%%%%%%%%%%%%%%%%%%%%%%%%%%%%%%%%%%%%%%%%%%%%%%%%%%%%%%%%%%%%%%%%%%%%%%%%%%%%%
\section{Introduction}
%%%%%%%%%%%%%%%%%%%%%%%%%%%%%%%%%%%%%%%%%%%%%%%%%%%%%%%%%%%%%%%%%%%%%%%%%%%%%%%%%%%

The discovery of the late-time cosmic accelerated expansion \cite{Perlmutter:1998np,Riess:1998cb} has motivated an extensive amount of research on modifications of general relativity (GR) \cite{modgrav,modgrav2,modgrav3,Capozziello:2011et,Lobo:2008sg}, as a possible cause of this cosmic speed-up. A plethora of theories have been proposed in the literature, essentially based on specific approaches. For instance, one may tackle the problem with the metric formalism, which consists on setting the Levi-Civita connection and varying the action with respect to the metric, or consider the metric-affine formalism \cite{Olmo:2011uz}, where the metric and the affine connection are regarded as independent variables. Note that the metric $g_{\mu\nu}$ may be thought of as a generalization of the gravitational potential and is used to define notions such as distances, volumes and angles. On the other hand, the affine connection $\Gamma^{\mu}{}_{\alpha\beta}$ defines parallel transport and covariant derivatives.

From a mathematical point of view (but inspired by the desire of obtaining a unified field theory) the first step in going beyond Riemannian geometry was taken by Weyl \cite{Weyl},  who extended the notion of parallel transport by considering the possibility that when vectors are transported along a closed path, their lengths, and not only their directions, change. The non-integrability of the length was used by Weyl to find a geometric interpretation for the electromagnetic field, as well as an elegant way to unify electromagnetism and gravitation. Weyl's theory was generalized by Dirac \cite{Dirac}, who proposed the existence of two metrics: the first is the unmeasurable metric $ds_E$, which changes as a result of the
transformations in the standards of length, and a second metric, which is measurable, and which is given by the conformally invariant atomic metric $ds_A$.
The two metrics are conformally related, so that  $f(x)ds_E = ds_A$, where $f(x)$ can be taken as any function  that transforms as $f(x)/\sigma (x)$ under the conformal transformation $g_{\mu \nu}\rightarrow  \sigma ^2 g_{\mu \nu}$. For an introduction to the Weyl-Dirac theory see \cite{Israelit}. The Weyl geometry can be immediately generalized to include torsion. The corresponding geometric model is called the Weyl-Cartan geometry, and it was studied extensively from both the physical and mathematical points of view \cite{Weyl-Cartan1, Weyl-Cartan2, Weyl-Cartan3,Weyl-Cartan4}. For a  review of the basic geometric properties and of the physical applications of the Riemann-Cartan and Weyl-Cartan geometries, we refer the reader to \cite{NoPeBe}.

It is a basic result in differential geometry that the general affine connection may always be decomposed into three independent components \cite{Hehl:1994ue,Ortin:2015hya}, namely,
\begin{equation}
	\label{Connection decomposition}
	\Gamma^{\lambda}{}_{\mu\nu} =
	\left\lbrace {}^{\lambda}_{\phantom{\alpha}\mu\nu} \right\rbrace +
	K^{\lambda}{}_{\mu\nu}+
	L^{\lambda}{}_{\mu\nu} \,,
\end{equation}
where the first term is the Levi-Civita connection of the metric $g_{\mu\nu}$, given by the standard definition
\begin{equation}
	\label{christoffel}
	\left\lbrace {}^{\lambda}_{\phantom{\alpha}\mu\nu} \right\rbrace \equiv \frac{1}{2} g^{\lambda \beta} \left( \partial_{\mu} g_{\beta\nu} + \partial_{\nu} g_{\beta\mu} - \partial_{\beta} g_{\mu\nu} \right) \,.
\end{equation}
The second term $K^{\lambda}_{\phantom{\alpha}\mu\nu}$ is the contortion:
\begin{equation}
	\label{Contortion}	
K^{\lambda}{}_{\mu\nu} \equiv \frac{1}{2} T^{\lambda}{}_{\mu\nu}+T_{(\mu}{}^{\lambda}{}_{\nu)} \,,
\end{equation}
with the torsion tensor defined as $T^{\lambda}{}_{\mu\nu}\equiv 2 \Gamma^{\lambda}{}_{[\mu\nu]}  $.
The third term is the disformation
\begin{equation}
	\label{Disformation}
	L^{\lambda}{}_{\mu\nu} \equiv \frac{1}{2} g^{\lambda \beta} \left( -Q_{\mu \beta\nu}-Q_{\nu \beta\mu}+Q_{\beta \mu \nu} \right)  \,,
\end{equation}
which is defined in terms of the nonmetricity tensor: $Q_{\rho \mu \nu} \equiv \nabla_{\rho} g_{\mu\nu}$.

Thus, by making assumptions on the affine connection, one is essentially specifying a metric-affine geometry \cite{Jarv:2018bgs}. For instance, the standard formulation of GR assumes a Levi-Civita connection, which implies vanishing torsion and nonmetricity, while its teleparallel equivalent (TEGR), uses the Weitzenb\"{o}ck connection, implying zero curvature and nonmetricity \cite{Maluf:2013gaa}. A gravitational model in a Weyl-Cartan spacetime, in which the Weitzenb\"{o}ck condition of the vanishing of the sum of the curvature and torsion scalar was considered in \cite{Haghani1}. A kinetic term for the torsion was also included in the gravitational action. The field equations of the model can be obtained from a Hilbert-Einstein type variational principle, and they lead to a complete description of the gravitational field in terms of two fields, the Weyl vector and the torsion, respectively, both defined in a curved background. The cosmological applications of the model were investigated for a particular choice of the free parameters in which the torsion vector is proportional to the Weyl vector. In particular, a de Sitter type late time evolution can be naturally obtained from the field equations of the model.
The Weitzenb\"{o}ck condition of the exact cancellation of curvature and torsion in a Weyl-Cartan geometry was imposed into the gravitational action via a Lagrange multiplier in \cite{Haghani2}. The dynamical variables are the spacetime metric, the Weyl vector and the torsion tensor, respectively. However, once the Weitzenb\"{o}ck condition is imposed on the Weyl-Cartan spacetime, the metric is not dynamical, and the gravitational dynamics and evolution is completely determined by the torsion tensor. The gravitational field equations can be obtained from a variational principle, and they explicitly depend on the Lagrange multiplier. The case of Riemann-Cartan spacetimes with zero nonmetricity that mimics the teleparallel theory of gravity was also considered.

A relatively unexplored territory consists in another equivalent formulation of GR, which is denoted the symmetric teleparallel equivalent of GR (STEGR). Here, one considers a vanishing curvature and torsion, and it is the nonmetricity tensor $Q$ that describes the gravitational interaction.
The STEGR was presented in the original brief paper \cite{Nester:1998mp}, where the authors emphasize that the formulation brings a new perspective to bear on GR, and the gravitational interaction effects, via the nonmetricity, present a character similar to the Newtonian force and are derived from a potential, namely, the metric. However, the formulation is geometric and covariant. The topic was further explored in \cite{Adak:2008gd}, where the STEGR was represented by the most general quadratic and parity conserving lagrangian with lagrange multipliers for vanishing torsion and curvature. It was shown that the considered lagrangian is equivalent to the Einstein-Hilbert lagrangian for certain values of the coupling coefficients. Furthermore, it was also shown that in the gravitational analogue of the Lorenz gauge \cite{Mol:2014ooa}, the field equations can be written as a system of Proca equations, which may be of interest in the study of propagation of gravitational-electromagnetic waves.

More recently, the symmetric teleparallel theories of gravity were analysed in \cite{BeltranJimenez:2017tkd}, where an exceptional class was discovered which is consistent with a vanishing affine connection. In fact, based on this remarkable property, a simpler geometrical formulation of GR was proposed that is oblivious to the affine spacetime structure, thus fundamentally depriving gravity from any inertial character. The resulting theory is described by the Einstein-Hilbert action purged from the boundary term and is more robustly underpinned by the spin-2 field theory. This construction also provides a novel starting point for modified gravity theories, and presents new and simple generalisations where analytical self-accelerating cosmological solutions arise naturally in the early and late-time universe. These topics were further explored in \cite{Conroy:2017yln}, where the linear perturbations in flat space were analysed, and in \cite{Koivisto:2018aip,BeltranJimenez:2018vdo}.

A generalization of STEGR was considered in \cite{Jarv:2018bgs}, where a nonminimal coupling of a scalar field to the nonmetricity invariant was introduced. The similarities and differences with analogous scalar-curvature and scalar-torsion theories were considered by discussing the field equations, role of connection, conformal transformations, relation to the $f(Q)$ theory, and cosmological applications.
This recent class of scalar-nonmetricity theories was extended by considering a five-parameter quadratic nonmetricity scalar and including a boundary term \cite{Runkla:2018xrv}. The equivalents for general relativity and ordinary (curvature based) scalar-tensor theories were also obtained as particular cases.
These nonminimal couplings motivate us to explore modifications of STEGR by considering a coupling between nonmetricity and the matter Lagrangian, much is the spirit of the case treated in \cite{Goenner,Koivisto:2005yk,Bertolami:2007gv}.
In fact, the nonminimal curvature-matter coupling and generalizations were extensively explored in the literature (see for instance \cite{Allemandi:2005qs,Nojiri:2004bi,Bertolami:2008zh,Harko:2010mv,Harko:2011kv,Haghani:2013oma,Odintsov:2013iba,Tamanini:2013aca,Carloni:2015bua,Barrientos:2018cnx,Wu:2018idg}), and we refer the reader to \cite{Harko:2014gwa,BookHarkoLobo} for recent reviews. The nonminimal torsion-matter coupling was also analysed in detail \cite{Harko:2014sja,Harko:2014aja,Feng:2015awr,Zhai:2018xqb,Lin:2016nvj} and a dynamical system analysis was developed in \cite{Carloni:2015lsa}.

Thus, the aim of the present paper is to present an extension of the symmetric teleparallel gravity, by introducing a new class of theories where the nonmetricity $Q$ is coupled nonminimally to the matter Lagrangian, in the framework of the metric-affine
formalism. This work is outlined in the following manner: In Section \ref{secII}, we present and motivate the symmetric teleparallel equivalent of general relativity (STEGR). In Section \ref{secIII}, we consider an extended STEGR, by coupling a general function of the nonmetricity to the matter Lagrangian. In Section \ref{secIV}, we consider some cosmological applications and we conclude in Section \ref{secV} with a summary and some perspectives.

%%%%%%%%%%%%%%%%%%%%%%%%%%%%%%%%%%%%%%%%%%%%%%%%%%%%%%%%%%%%%%%%%%%%%%%%%%%%%%%%%%%
\section{Covariant Einstein Lagrangian}\label{secII}
%%%%%%%%%%%%%%%%%%%%%%%%%%%%%%%%%%%%%%%%%%%%%%%%%%%%%%%%%%%%%%%%%%%%%%%%%%%%%%%%%%%

In 1916, Einstein wrote down \cite{Einstein} a simple Lagrangian formulation for his field
equations
\begin{equation}  \label{gg}
L_E = g^{\mu\nu}\Big(
\left\lbrace {}^{\alpha}_{\phantom{\alpha}\beta\mu} \right\rbrace
\left\lbrace {}^{\beta}_{\phantom{\alpha}\nu\alpha} \right\rbrace
-\left\lbrace {}^{\alpha}_{\phantom{\alpha}\beta\alpha} \right\rbrace
\left\lbrace {}^{\beta}_{\phantom{\alpha}\mu\nu} \right\rbrace %
\Big)\,,
\end{equation}
featuring the Levi-Civita connection written here as the Christoffel symbols of the metric defined in Eq. (\ref{christoffel}). The more standard Lagrangian formulation first discovered by Hilbert in
1915, given by the metric Ricci scalar $\mathcal{R}$, contains additional
terms which involve second derivatives of the metric. In fact, $\mathcal{R}%
=L_E + L_B$, where the boundary term (a total derivative) is
\begin{equation}  \label{bb}
L_B = g^{\alpha\mu}\mathcal{D}_\alpha
\left\lbrace {}^{\nu}_{\phantom{\alpha}\mu\nu} \right\rbrace
- g^{\mu\nu}\mathcal{D}_\alpha
\left\lbrace {}^{\alpha}_{\phantom{\alpha}\mu\nu} \right\rbrace\,,
\end{equation}
where the $\mathcal{D}_\alpha$ is the covariant derivative with the
connection (\ref{christoffel}). The reason why the higher-derivative
formulation has become the standard one is that the Lagrangian (\ref{gg}) is
not covariant.

This can be repaired by promoting the partial derivatives of the metric in (%
\ref{christoffel}) to covariant ones. We will therefore introduce an
independent ``Palatini connection'' $\Gamma^\alpha{}_{\mu\nu}$, with a
covariant derivative $\nabla_\alpha$, in order to define the tensor
\begin{equation} \label{disf}
L^\alpha{}_{\beta\gamma} = -\frac{1}{2}g^{\alpha\lambda}\left( \nabla_\gamma
g_{\beta\lambda} + \nabla_\beta g_{\lambda\gamma} - \nabla_\lambda
g_{\beta\gamma}\right)\,,
\end{equation}
which is nothing but the disformation (\ref{Disformation}) explicitly written. This way, the invariant
\begin{equation}  \label{q}
Q = -g^{\mu\nu}\left( L^\alpha{}_{\beta\mu}L^\beta{}_{\nu\alpha} - L^\alpha{}%
_{\beta\alpha}L^\beta{}_{\mu\nu}\right)\,,
\end{equation}
is by construction equivalent to (minus) the Einstein Lagrangian (\ref{gg}), when
the covariant derivative reduces to the partial one, i.e.,
\begin{equation}
\nabla_\alpha \overset{0}{=} \partial_\alpha\,, \quad Q \overset{0}{=} -L_E\,.
\end{equation}
This gauge choice, denoted with the $0$, was called the \textit{coincident
gauge}, and shown to be consistent in the symmetric teleparallel geometry \cite{BeltranJimenez:2017tkd}.

Though in this geometry the connection $\Gamma^\alpha{}_{\mu\nu}$ has neither
curvature nor torsion, the connection (\ref{christoffel}) and its curvature
still play their physical roles. Note that the Dirac Lagrangian, connected
with the $\Gamma^\alpha{}_{\mu\nu}$ in the symmetric teleparallel geometry,
filters out everything but the Christoffel symbols (\ref{christoffel}) from $\Gamma^\alpha{}_{\mu\nu} =
\left\lbrace {}^{\alpha}_{\phantom{\alpha}\mu\nu} \right\rbrace
+ L^\alpha{}_{\mu\nu} \overset{0}{=} 0$. The $Q$-formulation is thus a
pretty subtle improvement of GR, since (minimally coupled) fermions are
still connected metrically \cite{Koivisto:2018aip}, and whilst the pure gravity sector is now
trivially connected, effectively nothing changes but just the
higher-derivative boundary term $L_B$ disappears from the action.

%%%%%%%%%%%%%%%%%%%%%%%%%%%%%%%%%%%%%%%%%%%%%%%%%%%%%%%%%%%%%%%%%%%%%%%%%%%%%%%%%%%
\section{Matter couplings}\label{secIII}
%%%%%%%%%%%%%%%%%%%%%%%%%%%%%%%%%%%%%%%%%%%%%%%%%%%%%%%%%%%%%%%%%%%%%%%%%%%%%%%%%%%

%%%%%%%%%%%%%%%%%%%%%%%%%%%%%%%%%%%%%%%%%%%%%%%%%%%%%%%%%%%%%%%%%%%%%%%%%%%%%%%%%%%
\subsection{Action and field equations}
%%%%%%%%%%%%%%%%%%%%%%%%%%%%%%%%%%%%%%%%%%%%%%%%%%%%%%%%%%%%%%%%%%%%%%%%%%%%%%%%%%%

More substantial distinctions arise in generalisations of the $f(Q)$ gravitational theory. In this work, we consider the action defined by two functions, given by
\begin{equation}  \label{qqm}
S = \int {\mathrm{d}}^4 x \sqrt{-g}\left[\frac{1}{2} f_1(Q) + f_2(Q)L_M\right]\,,
\end{equation}
where $L_M$ is a Lagrangian function for the matter fields.

Nonminimal couplings with a function of the $\mathcal{R}$ have been
considered extensively \cite{Goenner,Koivisto:2005yk,Bertolami:2007gv},
since they predict very interesting phenomenology. Due to the
higher-derivative property of the scalar $\mathcal{R}$, however, these
theories are best considered as effective theories which might become
problematical at certain limits. As an example, for a density of a canonical
scalar field $\phi$, the nonminimal coupling of the form $f_2(\mathcal{R}%
)L_\phi$ introduces a kinetic term which does not fit into the viable Horndeski class. Let us point out, however, that such problems are likely to disappear when this coupling is formulated in the metric-affine approach because the field equations remain second order, see e.g. \cite{Olmo:2014sra}.

Therefore the proposal is to reconsider the nonminimal curvature couplings
in the framework of $Q$-gravity. Since the scalar invariant $Q$ in Eq. (\ref%
{q}) involves no higher derivatives, a coupling $f_2(Q)L_\phi$ results in
second order equations of motion. The motivation is to see whether the
subtle improvement of the geometrical formulation, when implemented in the
matter sector, would allow more universally consistent and viable
realisations of the nonminimal curvature-matter coupling theories.

We define the nonmetricity tensor and its two traces as follows:
\begin{equation}
Q_{\alpha \mu \nu }=\nabla _{\alpha }g_{\mu \nu }\,,\quad Q_{\alpha
}=Q_{\alpha }{}^{\mu }{}_{\mu }\,,\quad \tilde{Q}_{\alpha }=Q^{\mu
}{}_{\alpha \mu }\,.
\end{equation}%
It is also useful to introduce the superpotential
\begin{eqnarray}
4P^{\alpha }{}_{\mu \nu } &=& -Q^{\alpha }{}_{\mu \nu } + 2Q_{(\mu %
\phantom{\alpha}\nu )}^{\phantom{\mu}\alpha } - Q^{\alpha }g_{\mu \nu }  \notag
\\
&&-\tilde{Q}^{\alpha }g_{\mu \nu }-\delta _{(\mu }^{\alpha }Q_{\nu )}\,, \label{super}
\end{eqnarray}
which, by using Eq. (\ref{disf}), can also be written as
\begin{equation}
P^\alpha{}_{\mu\nu} = -\frac{1}{2}L^\alpha{}_{\mu\nu} + \frac{1}{4}\lp Q^\alpha - \tilde{Q}^\alpha\rp g_{\mu\nu} -
\frac{1}{4}\delta^\alpha_{(\mu}\tilde{Q}_{\nu)}\,.
\end{equation}
One can readily check that $Q=-Q_{\alpha \mu \nu }P^{\alpha \mu \nu
}$ (with our sign conventions that are the same as in Ref. \cite{BeltranJimenez:2017tkd}).
For notational simplicity, let us introduce the following definitions
\begin{equation}
f=f_{1}(Q)+2f_{2}(Q)L_{M}\,,\quad F=f_{1}^{\prime }(Q)+2f_{2}^{\prime
}(Q)L_{M}\,,  \label{f_F}
\end{equation}%
where primes ($'$) stand for derivatives of the functions with respect to $Q$. We also specify the following variations
\begin{eqnarray}
T_{\mu \nu } &=&-\frac{2}{\sqrt{-g}}\frac{\delta (\sqrt{-g}{L}_{M})}{\delta
g^{\mu \nu }}\,, \label{emt} \\
H_{\lambda }{}^{\mu \nu } &=&-\frac{1}{2}\frac{\delta (\sqrt{-g}{L}_{M})}{%
\delta \Gamma _{\phantom{\lambda}\mu \nu }^{\lambda }}\,,
\end{eqnarray}%
as the energy-momentum tensor and the hyper-momentum tensor density,
respectively.

Varying the action (\ref{qqm}) with respect to the metric, one obtains the
gravitational field equations given by
\begin{eqnarray}
&&\frac{2}{\sqrt{-g}}\nabla_\alpha\left(\sqrt{-g}FP^\alpha{}_{\mu\nu}\right)
+ \frac{1}{2}g_{\mu\nu} f_1  \notag \\
&& + F\left( P_{\mu\alpha\beta}Q_{\nu}{}^{\alpha\beta}
-2Q_{\alpha\beta\mu}P^{\alpha\beta}{}_\nu\right) = -f_2 T_{\mu\nu} \,.\label{efe}
\end{eqnarray}
When varying the action (\ref{qqm}) with respect to the connection, there are two possibilities to impose the symmetric
teleparallelism. We can either use the ``inertial variation'' \cite{Golovnev:2017dox}  by setting the connection in its pure-gauge form in the action, or we can consider a general connection in the action but supplement it with lagrange multipliers to eliminate the curvature and torsion \cite{BeltranJimenez:2018vdo}. Either way, we now obtain
\begin{equation}
\nabla_\mu\nabla_\nu \left(\sqrt{-g} F P^{\mu\nu}{}_\alpha - f_2 H_\alpha{}%
^{\mu\nu}\right) = 0\,.  \label{cfe}
\end{equation}

%%%%%%%%%%%%%%%%%%%%%%%%%%%%%%%%%%%%%%%%%%%%%%%%%%%%%%%%%%%%%%%%%%%%%%%%%%%%%%%%%%%
\subsection{The divergence of the energy-momentum tensor}
%%%%%%%%%%%%%%%%%%%%%%%%%%%%%%%%%%%%%%%%%%%%%%%%%%%%%%%%%%%%%%%%%%%%%%%%%%%%%%%%%%%

To begin with, let us note that in the symmetric teleparallel geometry, the nonmetricity tensor
satisfies the Bianchi identity
\be
\nabla_{[\alpha}Q_{\beta]\mu\nu}=0\,. \label{bianchi3}
\ee
We would like to deduce the energy-momentum conservation, i.e., the metric divergence of the tensor $T_{\mu\nu}$ as defined in (\ref{emt}).
We denote the purely Riemannian quantities with the curly symbols, and thus the metric covariant derivative with the symbol (\ref{christoffel}) is written with $\mathcal{D}_\alpha$.

To arrive at a useful form for the divergence of the energy-momentum tensor, as the first step, we raise one index in Eq. (\ref{efe}):
\ba
\frac{2}{\sqrt{-g}} \nabla_\alpha\left(\sqrt{-g}FP^{\alpha\mu}{}_\nu\right)
& + & \frac{1}{2}\delta^\mu_\nu f_1  \nn \\
 +  F P^{\mu\alpha\beta}Q_{\nu\alpha\beta}  & = &  -f_2 T^{\mu}{}_{\nu} \,,\label{efe2}
\ea
which has, in fact, simplified the equation.
Now recall that in symmetric teleparallel geometry the connection
is the sum of the metric piece (\ref{christoffel}) and the disformation (\ref{disf}). Therefore, for example, for an arbitrary vector $V^\alpha$, we have
\be \label{mtop}
\nabla_\mu V^\alpha = \mathcal{D}_\mu V^{\alpha} + L^\alpha{}_{\mu\beta}V^\beta\,.
\ee
Applying the same reasoning to a mixed-index tensor density $v^\mu{}_\nu$, taking into account that $L^\alpha{}_{\alpha\mu} = -\frac{1}{2}Q_\mu$, we have
\be \label{ptom}
\mathcal{D}_\mu v^\mu{}_\nu = \nabla_\mu  v^\mu{}_\nu + L^\alpha{}_{\mu\nu}v^\mu{}_\alpha\,.
\ee

Now we can consider $v^\mu{}_\nu =  \nabla_\alpha\left(\sqrt{-g}FP^{\alpha\mu}{}_\nu\right)$ such a mixed-index tensor density. Therefore,
we obtain for its metric divergence
\ba
\mathcal{D}_\mu \nabla_\alpha\left(\sqrt{-g}FP^{\alpha\mu}{}_\nu\right) =  \nabla_\alpha \nabla_\beta\lp f_2H_\nu{}^{\alpha\beta}\rp \nn \\
+ L^\alpha{}_{\mu\nu}\nabla_\beta \left(\sqrt{-g}FP^{\beta\mu}{}_\alpha\right),
\ea
where, for the first term, we have exploited the fact that in symmetric teleparallel geometry $[\nabla_\mu,\nabla_\nu]=0$, and then
used the connection equation of motion (\ref{cfe}). Taking the divergence of the full field equation (\ref{efe2}) and using the above result leads to
\ba
-\mathcal{D}_\mu\lp f_2 T^\mu{}_\nu\rp & -& \frac{2}{\sqrt{-g}}\nabla_\alpha \nabla_\beta\lp f_2H_\nu{}^{\alpha\beta}\rp \nn \\
& = & L^\lambda{}_{\mu\nu}( 2\nabla_\alpha - Q_\alpha) \lp FP^{\alpha\mu}{}_\lambda \rp
	\nn \\
&+& \frac{1}{2}\partial_\nu f_1
 +  \mathcal{D}_\mu \lp F P^{\mu\alpha\beta}Q_{\nu\alpha\beta}\rp\,. \label{div1}
\ea

We can first separate the $\nabla_\alpha F$-terms by simply using the Leibniz rule on the second and the third line. The two terms we get combine to zero,
\ba
2(\nabla_\beta F) P^{\beta\mu}{}_{\lambda}L^\lambda{}_{\mu\alpha} + (\mathcal{D}_\mu F) P^{\mu\nu\lambda}Q_{\alpha\nu\lambda}
	\nn \\
= (\nabla_\beta F )P^{\beta\mu\lambda}\lp 2L_{\lambda\mu\alpha}+Q_{\alpha\mu\lambda}\rp =0\,,
\ea
where in the first equality, we have just regrouped the terms, and in the second equality noted that $P^{\beta\mu\lambda}=P^{\beta(\mu\lambda)}$.
Thus, Eq. (\ref{div1}) takes the following form
\ba
-\mathcal{D}_\mu\lp f_2 T^\mu{}_\nu\rp  &-& \frac{2}{\sqrt{-g}}\nabla_\alpha \nabla_\beta\lp f_2H_\nu{}^{\alpha\beta}\rp
	\nn \\
 &=&  F\lp 2\nabla_\beta P^{\beta\mu}{}_\lambda - P^{\beta\mu}{}_\lambda Q_\beta\rp L^\lambda{}_{\mu\nu}
 	\nn \\
 	&+& \frac{1}{2}f_{1,\nu} +
F\mathcal{D}_\mu\lp P^{\mu\beta\lambda}Q_{\nu\beta\lambda}\rp\,.
\ea

Next, we rewrite the metric covariant derivative, in analogy with the expressions (\ref{mtop}) and (\ref{ptom}) and get
\ba
&-&\mathcal{D}_\mu\lp f_2 T^\mu{}_\nu\rp  - \frac{2}{\sqrt{-g}}\nabla_\alpha \nabla_\beta\lp f_2H_\nu{}^{\alpha\beta}\rp \nn \\
& = & F\lp 2\nabla_\beta P^{\beta\mu}{}_\lambda - P^{\beta\mu}{}_\lambda Q_\beta +  P^{\mu\alpha\beta}Q_{\lambda\alpha\beta} \rp L^\lambda{}_{\mu\nu} \nn \\ &+&
\frac{1}{2}\lp F  P^{\mu\alpha\beta}Q_\mu Q_{\nu\alpha\beta} +
f_{1,\nu}\rp +
F\nabla_\mu\lp P^{\mu\beta\lambda}Q_{\nu\beta\lambda}\rp\,. \label{div2}
\ea
We can then easily deal with the two derivative terms. By using again the symmetry $P^{\alpha\mu\nu} = P^{\alpha(\mu\nu)}$ and the identity (\ref{bianchi3}),
it is not difficult to see that
\ba
&& 2(\nabla_\beta P^{\beta\mu}{}_\lambda) L^\lambda{}_{\mu\nu} + \nabla_\mu\lp P^{\mu\beta\lambda}Q_{\nu\beta\lambda}\rp \nn \\
% & = & 2P^{\beta\mu\alpha}Q_{\beta\alpha\lambda}L^\lambda{}_{\mu\nu} + 2(\nabla_\beta P^{\beta\mu\alpha})L_{\alpha\mu\nu} \nn \\
%&+&  \nabla_\mu \lp P^{\mu\beta\lambda}Q_{\nu\beta\lambda}\rp \nn \\
%& = &  2P^{\beta\mu\alpha}Q_{\beta\alpha\lambda}L^\lambda{}_{\mu\nu} + (\nabla_\beta P^{\beta\mu\alpha})\lp 2L_{\alpha\mu\nu}+Q_{\nu\mu\alpha}\rp
%\nn \\ & + & P^{\mu\beta\lambda}\nabla_\mu Q_{\nu\beta\lambda} \nn \\
& = & 2P^{\beta\mu\alpha}Q_{\beta\alpha\lambda}L^\lambda{}_{\mu\nu} + P^{\mu\alpha\beta}\lp \nabla_\nu Q_{\mu\alpha\beta}\rp\,. \label{res1}
\ea
Furthermore, by a straightforward but tedious calculation using the definition (\ref{super}), one can show that
\ba
\lp \nabla_\alpha P^{\mu\nu\lambda}\rp Q_{\mu\nu\lambda}
& = &P^{\mu\nu\lambda}\lp\nabla_\alpha Q_{\mu\nu\lambda}\rp + \Big( 4P^{\beta\mu\gamma} Q_{\beta\gamma\lambda}
	\nn \\
&&+ 2P^{\mu\gamma\beta}Q_{\lambda\gamma\beta}- 2P^{\beta\mu}{}_\lambda Q_\beta\Big)L^\lambda{}_{\mu\alpha}
	\nn \\
&& -  2P^{\mu\nu\beta}Q_\mu Q_{\nu\alpha\beta}\,. \label{res2}
\ea
Using this information in Eq. (\ref{div2}), we then arrive at the final result as follows:
\ba
&-&\mathcal{D}_\mu\lp f_2 T^\mu{}_\nu\rp  - \frac{2}{\sqrt{-g}}\nabla_\alpha \nabla_\beta\lp f_2H_\nu{}^{\alpha\beta}\rp \nn \\
& = & F\lp 2P^{\beta\mu\gamma} Q_{\beta\gamma\lambda} - P^{\beta\mu}{}_\lambda Q_\beta +  P^{\mu\alpha\beta}Q_{\lambda\alpha\beta} \rp L^\lambda{}_{\mu\nu} \nn \\ &+&
\frac{1}{2}\lp F  P^{\mu\alpha\beta}Q_\mu Q_{\nu\alpha\beta} +
f_{,\alpha}\rp +
F P^{\mu\beta\lambda}\lp \nabla_\nu Q_{\mu\beta\lambda}\rp \nn \\
& = & \frac{1}{2}f_{1,\nu} + \frac{F}{2}\lb \lp\nabla_\nu P^{\mu\beta\lambda}\rp Q_{\mu\beta\lambda} +  P^{\mu\beta\lambda}\lp \nabla_\nu Q_{\mu\beta\lambda}\rp\rb \nn \\
& = &  \frac{1}{2}f_{1,\nu} - \frac{F}{2} Q_{,\nu} = - L_M f_{2,\nu}\,.
\ea
In the four steps above, we have substituted the results given by Eqs. (\ref{res1}) and (\ref{res2}), and then used the definitions of $Q$ and of $F$, respectively.

We may write this result explicitly as
\ba
\mathcal{D}_\mu T^\mu{}_\nu &+&   \frac{2}{\sqrt{-g}}\nabla_\alpha \nabla_\beta H_\nu{}^{\alpha\beta} \nn \\
& = & -\frac{2}{\sqrt{-g}f_2}\lb\lp\nabla_\alpha\nabla_\beta f_2\rp H_\nu{}^{\alpha\beta} + 2f_{2,(\alpha}\nabla_{\beta)}H_\nu{}^{\alpha\beta}\rb\nn \\
& - & \lp T^\mu{}_\nu - \delta^\mu_\nu L_M \rp \nabla_\mu \log{ f_2}\,. \label{divergence}
\ea
The second line is due to the nonminimal coupling of the hypermomentum, which, perhaps interestingly, can now contribute also directly and not only via its (second) derivatives. The third line is due to the nonminimal coupling of the energy-momentum tensor. This term is second order in derivatives (assuming of course that the $L_M$ does not contain higher derivatives), which confirms our optimistic expectation.

%%%%%%%%%%%%%%%%%%%%%%%%%%%%%%%%%%%%%%%%%%%%%%%%%%%%%%%%%%%%%%%%%%%%%%%%%%%%%%%%%%%%
\subsection{The energy and momentum balance equations}
%%%%%%%%%%%%%%%%%%%%%%%%%%%%%%%%%%%%%%%%%%%%%%%%%%%%%%%%%%%%%%%%%%%%%%%%%%%%%%%%%%%%

The expression of the divergence of the energy-momentum tensor, as given by Eq.~(\ref{divergence}), shows that due to the coupling between the nonmetricity $Q$ and the matter fields, in the present theory the matter energy-momentum tensor is no longer conserved. Generally, in theories with non-conserved divergence of the energy-momentum tensor we can write $\mathcal{D}_\mu {T}^\mu{}_\nu=A_{\nu}$, where $A_{\nu}$ is a model-dependent four-vector. In order to find a physical interpretation of $A_{\nu}$ we consider that the matter content of the gravitating system can be described by the energy-momentum tensor of a perfect fluid, given by
\be\label{35}
T_{\mu \nu}=\left(\rho+p\right)u_{\mu}u_{\nu}+pg_{\mu \nu},
\ee
where $\rho $ and $p$ are the thermodynamic energy and pressure, with the four-velocity $u_{\mu}$ satisfying the normalization condition $u_{\mu}u^{\mu}=-1$, and the differential identity $u^{\nu}\mathcal{D}^{\mu}u_{\nu}=0$, respectively. We also introduce the projection operator $h_{\lambda}^{\nu}=\delta _{\lambda}^{\nu}+u_{\lambda}u^{\nu}$, which satisfies the algebraic relation $u_{\nu}h^{\nu}_{\lambda}=0$.  By taking the divergence of the energy-momentum tensor given by Eq.~(\ref{35}), we obtain
\ba\label{36}
&&\left(\mathcal{D}^{\mu}\rho+\mathcal{D}^{\mu}p\right)u_{\mu}u_{\nu}+\left(\rho+p\right)u_{\nu}\mathcal{D}^{\mu}u_{\mu}
	\nonumber\\
&&+\left(\rho+p\right)u_{\mu}\mathcal{D}^{\mu}u_{\nu}+g_{\mu \nu}\mathcal{D}^{\mu}p=A_{\nu}.
\ea
We multiply first Eq.~(\ref{36}) by $u^{\nu}$, which immediately gives
\be\label{37}
\dot{\rho}+3\mathcal{H}\left(\rho+p\right)=A_{\nu}u^{\nu}=\mathcal{S},
\ee
where we have denoted the overdot as $\dot\; =u_{\mu}\mathcal{D}^{\mu}$, and considered the following definitions: $\mathcal{H}=(1/3)\mathcal{D}^{\mu}u_{\mu}$, and $\mathcal{S}=A_{\nu}u^{\nu}$, respectively. We multiply now Eq.~(\ref{36}) with the projection operator  $h_{\lambda}^{\nu}$, thus obtaining
\be
h^\nu_\lambda\left[ \left(\rho +p\right)\dot{u}_{\nu} + \mathcal{D}_\nu p\right]=h^{\nu}_{\lambda}A_{\nu}\,,
%\left(\rho +p\right)u^{\mu}\mathcal{D}_{\mu}u^{\lambda}=A_{\nu}h^{\nu}_{\lambda},
\ee
or, equivalently,
\be\label{39}
\frac{d^2x^{\lambda}}{ds^2}+\left\lbrace {}^{\lambda}_{\phantom{\alpha}\mu\nu} \right\rbrace  u^{\mu}u^{\nu}=\frac{h^{\lambda \nu}}{\rho+p}\left(A_{\nu}-\mathcal{D}_{\nu}p\right)=\mathcal{F}^{\lambda},
\ee
which translates as non-geodesic motion, where $\mathcal{F}^{\lambda}$ is an extra-force arising due to the $Q$-matter coupling.

Equation (\ref{37}) gives the energy balance equation in modified gravity with $Q$-couplings, or, in other words, the amount of energy entering or going out from a given volume. The term $\mathcal{S}$ acts as a source for the energy creation/annihilation. The matter energy of the gravitating system is conserved only if the condition $A_{\nu}u^{\nu}\equiv 0$ is satisfied in a given spacetime volume. If $A_{\nu}u^{\nu}\neq 0$, then particles or energy transfer processes must take place in the system. One such particular physical process that could be described by an energy  balance equation of type (\ref{37}) is represented by particle creation that could result from the irreversible energy transfer from the gravitational field to matter \cite{Harko:2014pqa,Harko:2015pma}. By taking into account the explicit form of the divergence of the energy-momentum tensor
%as given by Eq.~(\ref{22}), we obtain for the energy source term the expression
%\ba
%\mathcal{S}&=&  -\lp\log{f_2}\rp_{,\alpha}\rho u^{\alpha}- \frac{2}{f_2}\mathcal{D}_\mu \lp F_{,\alpha}P^{\alpha\mu}{}_\nu\rp u^{\nu}\nn \\
%&&- \frac{1}{2f_2}\lp f_1+FQ\rp_{,\nu}u^{\nu} + \frac{1}{f_2}F_{,\alpha}\mathcal{G}^\alpha{}_\nu u^{\nu}.
%\ea
%The first term in the above equation, proportional to $\rho u^{\alpha}$, gives the hydrodynamic flow of matter in or out the system. The other terms are of purely gravitational origin, and they may be associated to particle creation in the gravitational field.
from the result (\ref{divergence}), we can decompose the energy source term as
\begin{equation}
\mathcal{S}=\mathcal{S}_\mathcal{T} + \mathcal{S}_\mathcal{H}\,,
\end{equation}
where $\mathcal{S}_\mathcal{T}$ is defined by
\be \label{s_t}
\mathcal{S}_\mathcal{T} = \lp \rho + L_M \rp \frac{\dot{f_2}}{f_2}\,,
\ee
and the hypersource is given as
\begin{eqnarray}
\mathcal{S}_\mathcal{H} & = &  -\frac{2}{\sqrt{-g}}u^\nu\Big[\nabla_\alpha \nabla_\beta H_\nu{}^{\alpha\beta} \nonumber \\
& + & \frac{1}{f_2}\lp\nabla_\alpha\nabla_\beta f_2\rp H_\nu{}^{\alpha\beta} + \frac{1}{f_2}f_{2,(\alpha}\nabla_{\beta)}H_\nu{}^{\alpha\beta}\Big]\,.
\end{eqnarray}
Note that the energy source (\ref{s_t}) vanishes for perfect fluids, when we adopt the Lagrangian prescription $L_M=-\rho$.

Equation (\ref{39}) gives the equation of motion of massive particles in modified gravity with $Q$-coupling. From its general form it immediately follows that the motion is not geodesic, and an extra-force with components $\mathcal{F}^{\lambda}$ exerts a supplementary force on any particle. The extra force is orthogonal to the matter four-velocity, since, due to the presence of the projection operator in its expression, we always have $\mathcal{F}^{\lambda}u_{\lambda}\equiv 0$. This result points towards the fact that the extra-force as given by Eq.~(\ref{39}) is physical, since it satisfies the usual condition  for a ``normal'' force,  which requires that only the
components of the four-force that are orthogonal to the
four-velocity of the particle can influence its trajectory. In modified gravity with $Q$-couplings, the extra-force
%can be obtained easily as
%\ba\label{41}
%\mathcal{F}^{\lambda}&=&-\frac{h^{\alpha \lambda}}{\rho+p}\left[\left(\ln f_2\right)_{,\alpha}p+\mathcal{D}_{\alpha}p\right]
%	\nonumber\\
%&&+\frac{h^{\lambda \nu}}{\rho+p}\Bigg\{ - \frac{2}{f_2}\mathcal{D}_\mu \lp F_{,\alpha}P^{\alpha\mu}{}_\nu\rp
%\nonumber\\
%&&
%- \frac{1}{2f_2}\lp f_1+FQ\rp_{,\nu} + \frac{1}{f_2}F_{,\alpha}\mathcal{G}^\alpha{}_\nu\Bigg\} \ .
%\ea
%The extra-force given by Eq.~(\ref{41}) has the interesting property of being proportional to the Einstein tensor.
can be written, by recalling again the result (\ref{divergence}), as
\be
\mathcal{F}^{\lambda} = -\frac{h^{\alpha \lambda}\nabla_\alpha p}{\rho+p} + \mathcal{F}^{\lambda}_{\mathcal{T}} + \mathcal{F}^{\lambda}_{\mathcal{H}}\,,
\ee
where the first term on the right-hand-side is the usual general relativistic contribution of the pressure gradient, and the extra force consists of the following terms:
\be \label{f_t}
\mathcal{F}^{\lambda}_{\mathcal{T}}  = \left(-p+L_M\right)h^\lambda_\nu\nabla^\nu \log{f_2}\,,
\ee
and the hyperforce
\begin{eqnarray}
\mathcal{F}^\lambda_\mathcal{H} & = &  -\frac{2}{\sqrt{-g}}h^{\lambda\nu}\Big[\nabla_\alpha \nabla_\beta H_\nu{}^{\alpha\beta} \nonumber \\
& + & \frac{1}{f_2}\lp\nabla_\alpha\nabla_\beta f_2\rp H_\nu{}^{\alpha\beta} + \frac{1}{f_2}f_{2,(\alpha}\nabla_{\beta)}H_\nu{}^{\alpha\beta}\Big]\,,
\end{eqnarray}
respectively. It is interesting that the extra force (\ref{f_t}) vanishes identically for a perfect fluid if we adopt the Lagrangian prescription $L_M=p$, in which case the source term (\ref{s_t}) in turn would be non-vanishing.

%%%%%%%%%%%%%%%%%%%%%%%%%%%%%%%%%%%%%%%%%%%%%%%%%%%%%%%%%%%%%%%%%%%%%%%%%%%%%%%%%%%
\section{Cosmological application}\label{secIV}
%%%%%%%%%%%%%%%%%%%%%%%%%%%%%%%%%%%%%%%%%%%%%%%%%%%%%%%%%%%%%%%%%%%%%%%%%%%%%%%%%%%

We now explore several cosmological applications. For this purpose, consider the isotropic, homogeneous and spatially flat line element given by
\begin{equation}  \label{frw}
{\mathrm{d}} s^2 = -N^2(t){\mathrm{d}} t^2 + a^2(t)\delta_{ij} {\mathrm{d}}
x^i {\mathrm{d}} x^j\,,
\end{equation}
where we have included the lapse function $N(t)$  for generality, though in the present case we have the usual time reparametrization
freedom and may impose $N=1$ at any time. It is then convenient to define the expansion
and the dilation rates as
\begin{equation}
H=\frac{\dot{a}}{a}\,, \qquad T=\frac{\dot{N}}{N}\,,
\end{equation}
respectively. We shall work in the coincident gauge, and it is
straightforward to obtain that $Q=6(H/N)^2$.

We shall assume standard perfect fluid matter, whose energy-momentum tensor given by (\ref{35}) is diagonal. The field equations (\ref{efe}) in this case imply the following two
generalized Friedmann equations:
\begin{eqnarray}
f_2 \,\rho & = & \frac{f_1}{2} - 6F\frac{H^2}{N^2}\,,  \label{f1} \\
-f_2 \, p & = & \frac{f_1}{2} - \frac{2}{N^2}\left[ \left( \dot{F}-FT\right)
H + F\left( \dot{H}+3H^2\right)\right]\,, \,\,\,\,\,\,  \label{f2}
\end{eqnarray}
respectively. It is easy to check that in the limit of standard GR, $f_{1}=-Q$ and $f_{2}=1=-F$,
these reduce to the standard Friedmann equations. The equation of motion for
the connection (\ref{cfe}) is identically satisfied for the theory (\ref{qqm}%
) in the background (\ref{frw}). The continuity equation of matter can be
deduced from the above two equations (\ref{f1}) and (\ref{f2}), and is given
by
\begin{equation}
\dot{\rho} + 3H(\rho + p) = -\frac{6f_2'H}{f_2 N^2}\left( \dot{H} -HT\right) \left( L_M + \rho \right)\,.
\end{equation}
This is in accordance with the general result (\ref{s_t}).
Since in the minisuperspace given by Eq. (\ref{frw}) setting $L_M=-\rho$, we recover the standard
continuity equation
\begin{equation}
\dot{\rho}+3H(\rho +p)= 0 \,.  \label{cont1}
\end{equation}
This is compatible with the fact that the connection equation (\ref{cfe}) is trivialised in the isotropic and homogeneous
background.

%\begin{equation}
%\dot{\rho}+3H(\rho +p)=\frac{6f_{2}^{\prime }}{f_{2}^{2}N^{4}}\left(
%fN^{2}+12FH^{2}\right) H\left( \dot{H}-HT\right) \,.  \label{cont1}
%\end{equation}
%As expected, the standard conservation law is recovered when $f_{2}^{\prime
%}(Q)=0$, regardless of the function $f_{1}(Q)$. The non-conservation term is
%more precisely related to the time derivative of the function $f_{2}(Q)$,
%and we can write the Eq. (\ref{cont1}) more compactly
%\begin{equation}
%\dot{\rho}+3H(\rho +p)=-\frac{\dot{f_{2}}}{f_{2}^{2}}\left( f-2FQ\right) \,.
%	\label{cont2}
%\end{equation}

%%%%%%%%%%%%%%%%%%%%%%%%%%%%%%%%%%%%%%%%%%%%%%%%%%%%%%%%%%%%%%%%%%%%%%%%%%%%%%%%%%%
\subsection{The cosmological evolution equations}
%%%%%%%%%%%%%%%%%%%%%%%%%%%%%%%%%%%%%%%%%%%%%%%%%%%%%%%%%%%%%%%%%%%%%%%%%%%%%%%%%%%

In the following, we will adopt the gauge $N=1$, thus working in the
framework of standard Friedman-Robertson-Walker (FRW) geometry. With this choice we have
\begin{equation}  \label{21}
Q=6H^{2},
\end{equation}
and $T=0$, respectively. Therefore the field equations (\ref{f1}) and (\ref%
{f2}) can be reformulated as
\begin{equation}
3H^{2}=\frac{f_{2}}{2F}\left(- \rho +\frac{f_{1}}{2f_{2}}\right) ,  \label{f3}
\end{equation}
\begin{equation}
\dot{H}+3H^{2}+\frac{\dot{F}}{F}H=\frac{f_{2}}{2F}\left( p+\frac{f_{1}}{%
2f_{2}}\right) .  \label{f4}
\end{equation}
By eliminating the term $3H^{2}$ between the above two equations we obtain the following evolution equation for $H$
\begin{equation}  \label{f5}
\dot{H}+\frac{\dot{F}}{F}H=\frac{f_{2}}{2F}\left( \rho +p\right) .
\end{equation}

From Eq.~(\ref{f3}) we obtain the matter density as a function of $Q$ in the form
\be
\rho(Q)=\frac{\left(f_1/2f_2\right)\left[1-2\left(f_1'/f_1\right)Q\right]}{1-\left(f'_2/f_2\right)Q} \ .
\ee

After adding Eqs. (\ref{f4}) and (\ref{f5}), and by introducing the
effective energy density $\rho _{\rm eff}$ and effective pressure $p_{\rm eff}$ of
the cosmological fluid, defined as
\begin{equation}
\rho_{\rm eff}=-\frac{f_{2}}{2F}\left( \rho -\frac{f_{1}}{2f_{2}}\right) ,
\end{equation}
\begin{equation}
p_{\rm eff}=\frac{2\dot{F}}{F}H-\frac{f_{2}}{2F}\left( \rho +2p+\frac{f_{1}}{%
2f_{2}}\right) ,
\end{equation}%
we can write the gravitational field equations in a form similar to the
Friedmann equations of GR as
\begin{eqnarray}
3H^{2}&=&\rho_{\rm eff},  \label{f6}
	\\
%\end{equation}
%\begin{equation}
2\dot{H}+3H^{2}&=&-p_{\rm eff}.  \label{f7}
\end{eqnarray}

An important cosmological quantity, the deceleration parameter, defined as
\begin{equation}
q=\frac{d}{dt}\frac{1}{H}-1=-\frac{\dot{H}}{H^{2}}-1,
\end{equation}%
can be obtained from Eq. (\ref{f5}) as
\begin{equation}
q=\frac{\dot{F}}{F}\frac{1}{H}-\frac{f_{2}}{2H^{2}F}\left( \rho +p\right) -1.
\label{dec}
\end{equation}
Moreover, to describe cosmological evolution, and the possible transition to
an accelerated phase, we also introduce the parameter $w$ of the dark energy
equation of state, defined as
\begin{equation}
w=\frac{p_{\rm eff}}{\rho_{\rm eff}}=\frac{-4\dot{F}H+f_{2}\left( \rho +2p+\frac{f_{1}%
}{2f_{2}}\right) }{f_{2}\left( \rho -\frac{f_{1}}{2f_{2}}\right) }.
\label{w}
\end{equation}
Alternatively, the deceleration parameter can be written as
\begin{equation}
q=\frac{1}{2}\left( 1+3w\right) = 2 + \frac{3\lp 4\dot{F}H - f_1- 2f_2 p \rp}{f_1-2f_2\rho}\,.
\end{equation}
%\begin{equation}
%q=\frac{1}{2}\left( 1+3w\right) =\frac{1}{2}\left[ 1+\frac{4\dot{F}%
%+f_{2}\left( \rho +2p+\frac{f_{1}}{2f_{2}}\right) }{f_{2}\left( \rho -\frac{%
%f_{1}}{2f_{2}}\right) }\right] .
%\end{equation}

%%%%%%%%%%%%%%%%%%%%%%%%%%%%%%%%%%%%%%%%%%%%%%%%%%%%%%%%%%%%%%%%%%%%%%%%%%%%%%%%%%%
\subsection{The de Sitter solution}
%%%%%%%%%%%%%%%%%%%%%%%%%%%%%%%%%%%%%%%%%%%%%%%%%%%%%%%%%%%%%%%%%%%%%%%%%%%%%%%%%%%

As a first step in considering explicit theoretical models, we consider the
problem of the existence of a de Sitter type vacuum solution of the
cosmological field equations. The de Sitter solution corresponds to $\rho
=p=0$, and $H=H_{0}=\mathrm{constant}$, respectively. For a vacuum de Sitter
type Universe, Eq. (\ref{f5}) immediately gives $\dot{F}=0$, and $F=\mathrm{%
constant}=F_{0}$. For the vacuum state $L_{M}=0$, and therefore the
definitions of $f$ and $F$ reduce to $f=f_{1}\left( Q\right) $, and $%
F=f_{1}^{\prime }\left( Q\right) $.

The condition $F=\mathrm{constant}=-F_{0}$ is satisfied for any $Q$ only in the case
\begin{equation}
f_{1}\left( Q\right) =-F_{0}Q-2\Lambda=-6F_{0}H_{0}^{2}-2\Lambda, \label{gr}
\end{equation}
where $\Lambda$ is an arbitrary constant of integration. In the vacuum de
Sitter phase both field equations (\ref{f6}) and (\ref{f7}) reduce to the
algebraic form
\begin{equation}
3H_{0}^{2}=\frac{6F_{0}H_{0}^{2}+2\Lambda}{4F_{0}},
\end{equation}%
or equivalently
\begin{equation}
H_{0}=\sqrt{\frac{\Lambda}{3F_{0}}}.
\end{equation}
The specific case of Eq. (\ref{gr}) is of course equivalent to GR with a cosmological constant $\Lambda$ with the
normalisation $F_0=1$.
However, there exist vacuum de Sitter solutions in very generic cases. The combination of the
field equations (\ref{f6}) and (\ref{f7}) supports consistently such solutions as long as
\be
\lp \log{f_1(Q)}\rp' = \frac{1}{12H_0^2}\,,
\ee
when the right hand side is evaluated at $Q=6H_0^2$.
As one can see immediately from Eqs. (\ref{dec}) and (\ref{w}), for the de
Sitter evolution we obtain $q=-1$, and $w=-1$, respectively.

%%%%%%%%%%%%%%%%%%%%%%%%%%%%%%%%%%%%%%%%%%%%%%%%%%%%%%%%%%%%%%%%%%%%%%%%%%%%%%%%%%%
\subsection{Cosmological models with specific forms of $f_{1}$ and $f_{2}$}
%%%%%%%%%%%%%%%%%%%%%%%%%%%%%%%%%%%%%%%%%%%%%%%%%%%%%%%%%%%%%%%%%%%%%%%%%%%%%%%%%%%

In order to investigate more general cosmological models, we need to fix the
functional form of the functions $f_1(Q)$ and $f_2(Q)$. Once this form is
fixed a priori, the system of gravitational equations becomes closed, and
their solutions can give a full description of the cosmological evolution.

%%%%%%%%%%%%%%%%%%%%%%%%%%%%%%%%%%%%%%%%%%%%%%%%%%%%%%%%%%%%%%%%%%%%%%%%%%%%%%%%%%%
\subsubsection{Power-law dependence of the nonminimal couplings}
%%%%%%%%%%%%%%%%%%%%%%%%%%%%%%%%%%%%%%%%%%%%%%%%%%%%%%%%%%%%%%%%%%%%%%%%%%%%%%%%%%%

As a first example of cosmological models of this type we consider the case
in which both $f_1$ and $f_2$ have a simple power-law dependence on $Q$, so
that
\begin{equation}
f_1(Q)=AQ^{\alpha +1}, \qquad f_2(Q)=BQ^{\beta +1},
\end{equation}
where $A$, $B$, $\alpha $ and $\beta $ are arbitrary constants. For the
matter Lagrangian we will adopt the expression $L_M=-\rho$. Moreover, we assume that the cosmological matter satisfies the linear barotropic equation of state with $p=(\gamma -1)\rho$. For the function $F$ we obtain the expression
\begin{eqnarray}
F(Q)&=&A(1+\alpha)Q^{\alpha}-2 B\left(1+\beta\right)Q^{\beta}%
\rho  \nn \\
&=&A(1+\alpha)\left(6H^2\right)^{\alpha}-2B\left(1+\beta\right)%
\left(6H^2\right)^{\beta}\rho.
\end{eqnarray}

By substituting the above expressions of $f_1$, $f_2$ and $F$ into Eq.~(\ref%
{f3}) allows us to obtain the energy density as
\begin{equation}  \label{38}
\rho= \frac{A(1+2\alpha) (6H^2)^{\alpha -\beta}}{B(2+4\beta)}.
\end{equation}
%\begin{equation}  \label{38}
%\rho(t)= \frac{(\alpha +2) 6^{\alpha -\beta }A H^{2 (\alpha -\beta )}(t)}{%
%(\beta +3) B}.
%\end{equation}
Hence, the evolution equation of the Hubble function (\ref{f5}) takes the
form
\begin{eqnarray}
\dot{H} = -\frac{3\gamma H^2}{2(\alpha-\beta)}\,,
\end{eqnarray}
which provides the following solution
\be
H(t)= \frac{2 H_0 (\alpha -\beta )}{2 (\alpha - \beta) +3 \gamma
   H_0 \left(t-t_0\right)},
\ee
where $H_0=H\left(t_0\right)$ and
\be
a(t)=a_0\left[2 (\alpha - \beta) +3 \gamma  H_0 \left(t-t_0\right)\right]^{\frac{2 (\alpha -\beta )}{3 \gamma
   }} \ ,
\ee
respectively.
This means that the Universe expands as if it was dominated by a fluid with the
effective equation of state parameter
\begin{equation}
\gamma_{eff} = \frac{\gamma}{\alpha-\beta}\,.
\end{equation}
%For $\alpha \neq -2$, Eq.~(\ref{39}) has the general solution
%\begin{equation}
%H(t)=\frac{H_{0}f(\alpha,\beta) }{3(\alpha +2)\gamma
%H_{0}(t-t_{0})+f(\alpha,\beta)},
%\end{equation}
%where for notational simplicity, the factor $f(\alpha,\beta)$ is defined by
%\begin{equation}
%f(\alpha,\beta)=2\alpha ^{2}(\beta +3)-2\alpha \left( \beta
%^{2}-4\right) -\beta (4\beta +5)+1,
%\end{equation}
%where we have used the initial condition $H\left( t_{0}\right)
%=H_{0}$. The evolution of the scale factor is given by
%\begin{equation}
%a(t)=\left[ 3(\alpha +2)\gamma H_{0}(t-t_{0})+f(\alpha,\beta)\right] ^{\frac{f(\alpha,\beta)}{3(\alpha +2)\gamma }},
%\end{equation}%
%(for $\alpha \neq -2$). The matter energy density varies in time as
%\begin{eqnarray}
%& \rho (t)=\frac{(\alpha +2)6^{\alpha -\beta }A}{(\beta +3)B}\left[
%\frac{H_{0}f(\alpha,\beta) }{3(\alpha +2)\gamma
%H_{0}(t-t_{0})+f(\alpha,\beta)}
%\right] ^{2(\alpha -\beta )}.
%\end{eqnarray}
%As for the deceleration parameter, it takes the simple, time independent
%form,
%\begin{equation}
%q=\frac{3 (\alpha +2) \gamma }{f(\alpha,\beta)}-1,
%\end{equation}
%(for $\alpha \neq -2$).

In this model the deceleration parameter has a constant value $q=3 \gamma /2 (\alpha -\beta )$ during the
entire cosmological evolution (when the equation
of state $\gamma$ of the cosmological matter is a constant). Depending on the numerical values of
$\alpha $ and $\beta$, a large range of cosmological behaviors can be
obtained, including both accelerating and decelerating phases, with the
possibility of the deceleration parameter of taking a $q\approx -1$ value.
In this case we obtain a power law type accelerating expansion of the
Universe. Exact de Sitter type evolution can however only be realised with
a cosmological constant $\gamma=0$.

%%%%%%%%%%%%%%%%%%%%%%%%%%%%%%%%%%%%%%%%%%%%%%%%%%%%%%%%%%%%%%%%%%%%%%%%%%%%%%%%%%%
\subsubsection{Exponential dependence of the nonminimal couplings}
%%%%%%%%%%%%%%%%%%%%%%%%%%%%%%%%%%%%%%%%%%%%%%%%%%%%%%%%%%%%%%%%%%%%%%%%%%%%%%%%%%%

As a second example of a cosmological scenario in the framework of the
matter-$Q$ field coupling theory we consider the case of the exponential
dependencies of the functions $f_{1}$ and $f_{2}$ on the $Q$-field, so that
\begin{equation}
f_{1}=Ae^{\alpha Q}, \qquad f_{2}=Be^{\beta Q},
\end{equation}%
where $A$, $B$, $\alpha $ and $\beta $ are, once again, arbitrary constants. For the
function $F$ we easily obtain
\begin{equation}
F(Q)=A\alpha e^{\alpha Q}-2\beta Be^{\beta Q}\rho .
\end{equation}%

In the following, we assume again that $Q=6H^{2}>0$, and $H(Q)=\sqrt{Q/6}$, $%
\dot{H}=\dot{Q}/2\sqrt{6}\sqrt{Q}$. Then from Eq.~(\ref{f3}), we obtain the
density of the matter as a function of $Q$ in the form
\begin{equation}
\rho (Q)=\frac{A \left[2 \alpha  Q(t)-1\right] e^{(\alpha -\beta ) Q(t)}}{2 B \left[2\beta  Q(t)-1\right]}.
\end{equation}%
By using the above representation of the density we obtain for the function $%
F$ the expression
\begin{equation}
F=\frac{A (\alpha-\beta ) e^{\alpha  Q(t)}}{1-2\beta Q(t)}\,.
\end{equation}
%\begin{equation}
%F=\frac{A (\beta -2 \alpha ) e^{\alpha  Q(t)}}{2\left[ \beta  Q(t)-1\right]}.
%\end{equation}

The evolution of the Hubble function (\ref{f5}) can be obtained, in
terms of $Q$, as the solution of the following first order differential equation
\begin{equation}  \label{49}
\frac{dQ}{dt}= \sqrt{\frac{3}{2}}\frac{ \gamma  \sqrt{Q} (1-2 \alpha  Q) (1-2
   \beta  Q)}{(\alpha -\beta ) \left(1+2 (\alpha +\beta )
   Q-4 \alpha  \beta  Q^2\right)}.
\end{equation}

%
%\begin{equation}  \label{49}
%\frac{dQ}{dt}= -\frac{\sqrt{6} \sqrt{Q(t)} \left[2 \alpha  Q(t)-1\right] \left[\beta  Q(t)-1\right]}{(2
%   \alpha -\beta ) \left[2 \alpha  \beta  Q(t)^2-2 \alpha  Q(t)-\beta
%   Q(t)-1\right]}.
%\end{equation}%
%\paragraph{Time evolution of the Universe}
The general solution of Eq.~(\ref{49}) is given by
\begin{eqnarray}
&&t(Q)-t_0=
\frac{2}{\gamma}\sqrt{\frac{2}{3}}
 \Bigg\{
   \sqrt{2\alpha } \Big[\tanh ^{-1}\left(\sqrt{2\alpha Q}\right)
   \nonumber\\
&&  - \tanh
   ^{-1}\left(\sqrt{2\alpha Q_0}\right)\Big]+
   \sqrt{2\beta }
   \Big[\tanh ^{-1}\left( \sqrt{2\beta Q_0}\right)
   \nonumber\\
&&  - \tanh
   ^{-1}\left(\sqrt{2\beta Q}\right)\Big]-(\alpha -\beta )\left(\sqrt{Q}-\sqrt{Q_0}\right)\Bigg\},
\end{eqnarray}
where we have used the initial condition $Q\left( t_{0}\right) =Q_{0}$.
Hence, we have obtained the general solution of the field equations in a
parametric form, with $Q$ taken as the parameter.

The evolution of the scale
factor can be obtained from the equation
\begin{equation}
\frac{1}{a}\frac{da}{dQ}=\sqrt{\frac{Q}{6}}\frac{dt}{dQ},
\end{equation}%
and is given by
\begin{equation}
a(Q)=a_{0}\frac{\left(1-2\beta Q\right)^2}{\left(1-2\alpha Q\right)^2}e^{-\left(\alpha -\beta \right)Q/3\gamma},
\end{equation}
where $a_0$ is an arbitrary constant of integration.

The deceleration parameter can be obtained as
\begin{equation}
q(Q)=-\frac{3}{2}\frac{ \gamma   (1-2 \alpha  Q) (1-2
	\beta  Q)}{(\alpha -\beta ) Q\left(1+2 (\alpha +\beta )
	Q-4 \alpha  \beta  Q^2\right)}-1.
\end{equation}

\begin{figure*}[htp!]
	\centering
	\includegraphics[width=8.5cm]{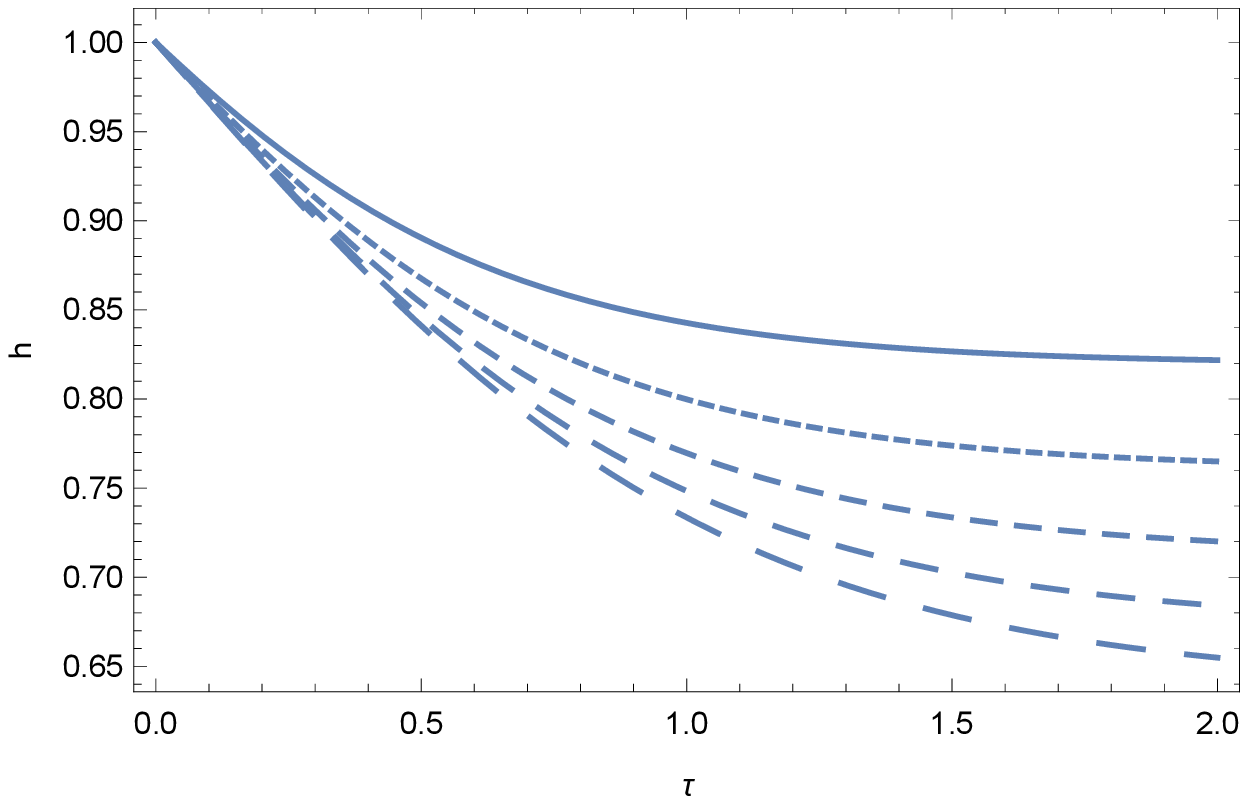} %
	\includegraphics[width=8.5cm]{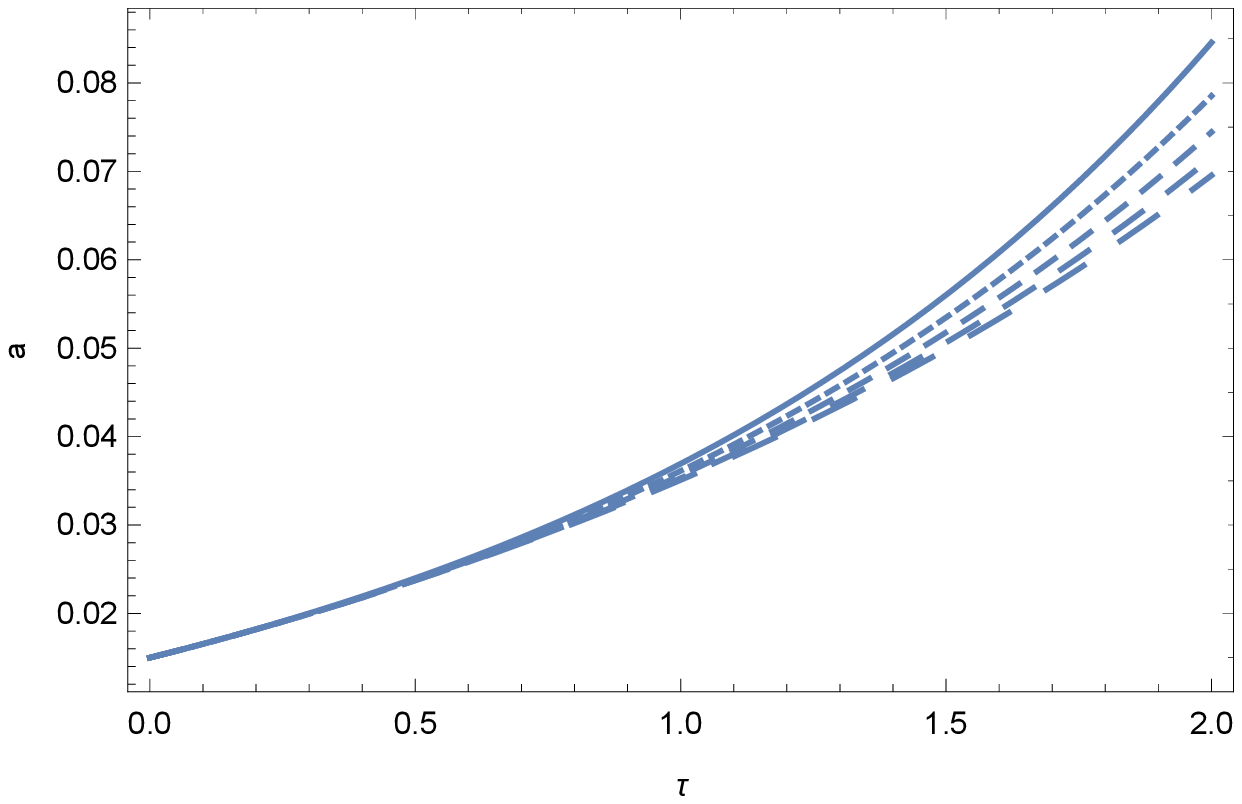}
	\caption{Specific case of the exponential dependence of the nonminimal couplings. Variation as function of the dimensionless time $\tau$ of the Hubble function $h$ (left figure) and of the scale factor $a$ (right figure) for a pressureless Universe with $\gamma =1$ for $\tilde{\beta}=0.054$, and different values of $\tilde{\alpha}$: $\tilde{\alpha}=0.124$ (solid curve), $\tilde{\alpha}=0.144$ (dotted curve), $\tilde{\alpha}=0.164$ (short dashed curve), $\tilde{\alpha}=0.184$ (dashed curve), and $\tilde{\alpha}=0.204$ (long dashed curve). For $Q_0$ we have adopted the initial value $Q_0=6$, and $a(0)=0.015$. We refer the reader to the text for more details.}
	\label{fig1}
\end{figure*}
\begin{figure*}[htp!]
	\centering
	\includegraphics[width=8.5cm]{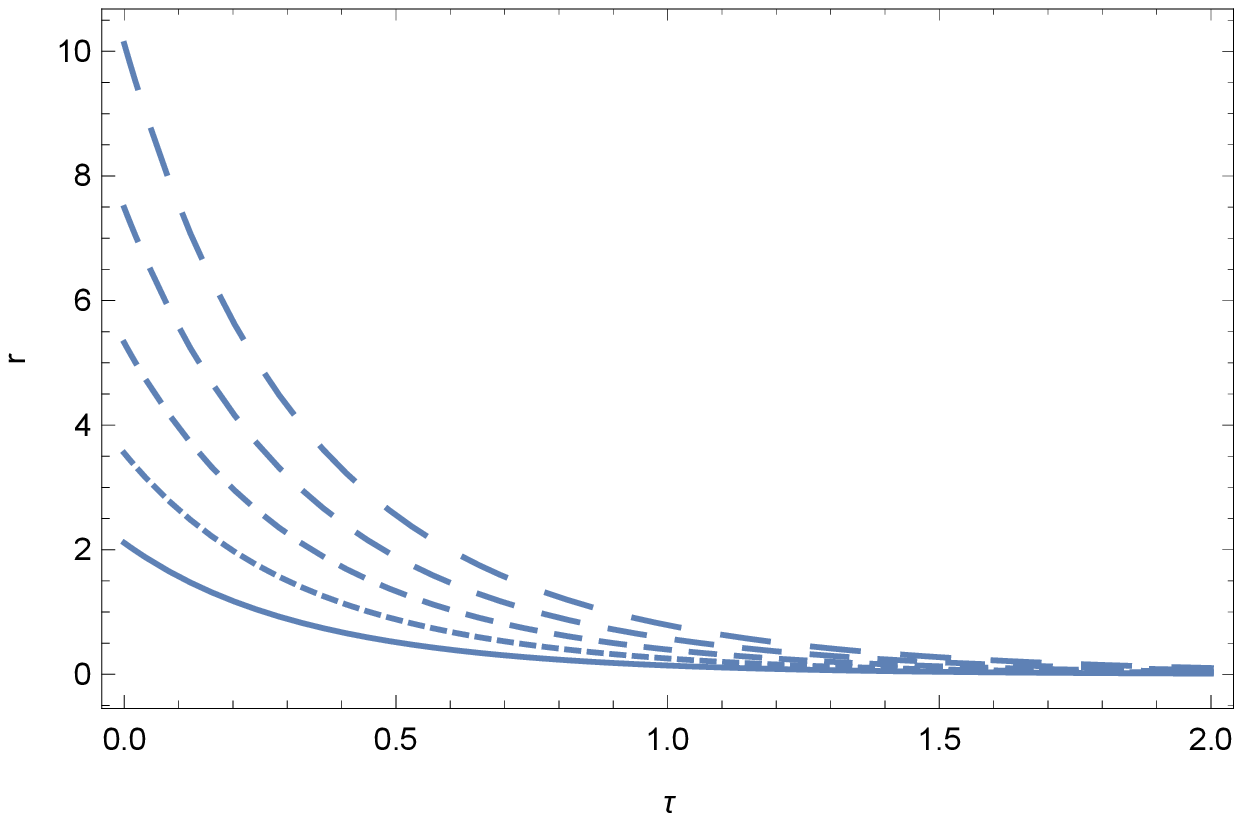} %
	\includegraphics[width=8.5cm]{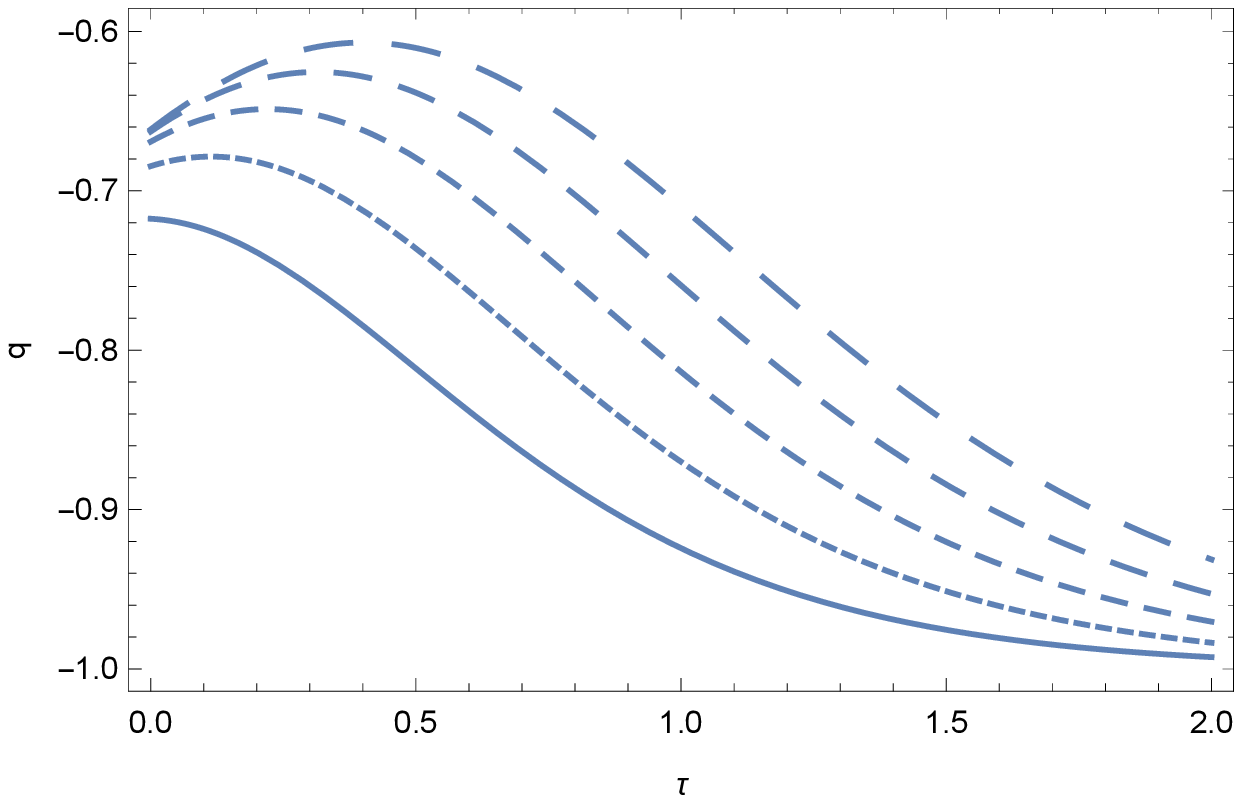}
	\caption{Specific case of the exponential dependence of the nonminimal couplings. Variation as function of the dimensionless time $\tau$ of the energy density of the matter $r$ (left figure) and of the deceleration parameter $q$ (right figure) for a pressureless Universe with $\gamma =1$ for $\tilde{\beta}=0.054$, and different values of $\tilde{\alpha}$: $\tilde{\alpha}=0.124$ (solid curve), $\tilde{\alpha}=0.144$ (dotted curve), $\tilde{\alpha}=0.164$ (short dashed curve), $\tilde{\alpha}=0.184$ (dashed curve), and $\tilde{\alpha}=0.204$ (long dashed curve). For $Q_0$ we have adopted the initial value $Q_0=6$, and $a(0)=0.015$. See the text for more details.}
	\label{fig2}
\end{figure*}

In order to obtain a dimensionless form of the cosmological evolution equations we introduce a set of dimensionless variables $\left(h, \tau, r, \tilde{\alpha},\tilde{\beta}, \tilde{Q}\right)$, defined as
\ba
H&=&H_0h, \qquad t=\frac{\tau}{H_0}, \qquad \rho=3H_0^2r,\nonumber\\
\alpha &=&\frac{\tilde{\alpha}}{H_0^2}, \qquad \beta =\frac{\tilde{\beta }}{H_0^2}, \qquad  Q=H_0^2\tilde{Q},
\ea
where $H_0$ is a fixed value of the Hubble function, which may correspond, for example, to the end of inflation, or to the present age of the Universe. For the ratio of the constants $A$ and $B$ we obtain the expression
\be
\frac{A}{B}=\frac{6H_0^2\left(2\beta Q_0 -1\right)}{2\left(2\alpha Q_0 -1\right)}e^{-\left(\alpha -\beta\right)Q_0},
\ee
where $Q_0=Q\left(\tau _0\right)$.
For $A/B>0$ the condition of the positivity of the matter energy density imposes the constraints $\alpha >1/12$ and $\beta >1/6$ on the model parameters $\alpha $ and $\beta $. All the dimensionless expressions of the time evolution, Hubble function, energy density and deceleration parameter can be simply obtained from the dimensional form by simply substituting the initial variables with the dimensionless ones. Hence we will not write down the explicit form of the dimensionless representation of the basic cosmological evolution equations,  and of their solutions.

The variations of the Hubble function, scale factor, matter energy density, and deceleration parameter are represented, for different values of the model parameters $\tilde{\alpha}$ and $\tilde{\beta}$ in Figs.~\ref{fig1} and \ref{fig2}, respectively.

As one can see from Fig.~\ref{fig1} the Hubble function is a monotonically decreasing function of the time, indicating an expansionary evolution of the Universe. In the large time limit  the rate of time variation $h$ is slow, and it shows a significant dependence on the numerical values of the model parameters $\tilde{\alpha}$ and $\tilde{\beta}$. The scale factor is a monotonically increasing function of time, and in the late stages of the cosmological evolution it shows a relatively weak dependence on $\tilde{\alpha}$ and $\tilde{\beta}$. The energy density of the matter, presented in the left panel of Fig.~\ref{fig2}, monotonically decreases in time, and in the large time limit it tends to zero in a way almost independent on $\tilde{\alpha}$ and $\tilde{\beta}$. However, the early time evolution is significantly influenced by the model parameters.  The evolution of the Universe begins in an accelerating state, with the deceleration parameter $q$, shown in the right panel of Fig.~\ref{fig2}, taking negative initial values of the order of $q\approx -0.70$. Then the Universe begins to accelerate, with $q$, showing a complex dynamics, decreasing in time. In the large time limit the Universe reaches the exponentially accelerating de Sitter phase with $q=-1$, a result which is independent on the model parameters.

%%%%%%%%%%%%%%%%%%%%%%%%%%%%%%%%%%%%%%%%%%%%%%%%%%%%%%%%%%%%%%%%%%%%%%%%%%%%%%%%%%%
\section{Summary and future outlook}\label{secV}
%%%%%%%%%%%%%%%%%%%%%%%%%%%%%%%%%%%%%%%%%%%%%%%%%%%%%%%%%%%%%%%%%%%%%%%%%%%%%%%%%%%

In this work, we have explored an extension of the symmetric teleparallel gravity, by considering a new class of theories where the nonmetricity $Q$ is coupled nonminimally to the matter Lagrangian, in the framework of the metric-affine
formalism. As in the standard curvature-matter couplings, this nonminimal $Q$-matter coupling entails the nonconservation of the energy-momentum tensor, and consequently the appearance of an extra force. We have verified whether the subtle improvement of the geometrical formulation, when implemented in the matter sector, would allow more universally consistent and viable realisations of the nonminimal curvature-matter coupling theories. Furthermore, we have also analysed several cosmological applications.

As a first step in this direction we have obtained the generalized Friedmann equations describing the cosmological evolution in flat FRW type geometry. The coupling between matter and the $Q$ field introduces two types of corrections. The first is the presence of a term of the form $f_2/2F$ multiplying the components of the energy-momentum tensor (energy density and pressure) in both Friedmann equations. Secondly, an additive term of the form $f_1/4F$ also appears in the generalized Friedmann equations. The basic equations describing the cosmological dynamics can then be reformulated in terms of an effective energy density and pressure, which both depend on the standard components of the energy-momentum tensor, and on the functions $f_i(Q)$, $i=1,2$, and on $F(Q,\rho)$.
In the vacuum case $\rho =p=0$, the deceleration parameter takes the form $q=-1+12\dot{F}H/f_1$, showing that, depending on the mathematical forms of the coupling functions, a large number of cosmological evolutionary scenarios can be obtained. Generally, we have shown explicitly that for late times, the Universe attains an exponentially accelerating de Sitter phase.

We have also considered two explicit classes of cosmological models obtained by choosing some specific functional forms for the functions $f_1(Q)$ and $f_2(Q)$, corresponding to power law and exponential forms of the couplings. In the case of the power law dependence of $f_i(Q)$, $i=1,2$, the field equations can be solved exactly, leading to a power-law dependence of the scale factor. The deceleration coefficient is constant, but by an appropriate choice of the parameters accelerating evolutions can be easily obtained. In the case of the exponential dependence of the couplings the overall cosmological dynamics of the Universe is very complex, and the relevant results can be obtained only by numerically integrating the evolution equation. The results are strongly dependent on the numerical values of the model parameters. For the specific range of cosmological parameters we have considered that the Universe is born in an accelerated phase, and in the large time limit it reaches the de Sitter phase, which acts as an attractor for the generalized Friedmann equations. This solution may represent an alternative to the standard inflationary scenario \cite{rev3,rev4}, in which the de Sitter phase is triggered by the presence of some cosmological scalar fields. Of course the present model is also valid when instead of ordinary matter one considers scalar fields. Considering inflation in $Q$-coupling gravity in the presence of scalar fields may give a new perspective on the physical, geometrical and cosmological processes that may have played a dominant role in the very early evolution of the Universe.

Thus, in summary, we have established the theoretical consistency and motivations on these extensions of $f(Q)$ family of theories. Furthermore, we considered cosmological applications, in which the presented approach provides gravitational alternatives to dark energy. As future avenues of research, one should aim in characterizing the phenomenology predicted by these theories with a nonmetricity-matter coupling, in order to find constraints arising from observations. The study of these phenomena may also provide some specific signatures and effects, which could distinguish and discriminate between the various theories of modified gravity. We also propose to use a background metric to analyse the dynamic system for specific nonmetricity-matter coupling models, and use the data of SNIa, BAO, CMB shift parameter to obtain restrictions for the respective models, and explore in detail the analysis of structure formation. Another topic that needs to be addressed is the analysis of the post-Newtonian formalism applied to this nonminimal extension of $f(Q)$ gravity, in order to pass the local gravity constraints. Work along these lines are presently underway, and are to be presented in the near future.

%%%%%%%%%%%%%%%%%%%%%%%%%%%%%%%%%%%%%%%%%%%%%%%%%%%%%%%%%%%%%%%%%%%%%%%%%%%%%%%%%\
\section*{Acknowledgments}

FSNL is funded by the Funda\c{c}\~ao para a Ci\^encia e a Tecnologia (FCT, Portugal) through an Investigador FCT Research contract No.~IF/00859/2012. GJO is funded by the Ramon y Cajal contract RYC-2013-13019 (Spain). DRG is funded by the FCT postdoctoral fellowship No.~SFRH/BPD/102958/2014. FSNL and DRG also acknowledge funding from the research grants UID/FIS/04434/2013 and No.~PEst-OE/FIS/UI2751/2014. This work is supported by the Spanish projects FIS2014-57387-C3-1-P,  FIS2017-84440-C2-1-P (AEI/FEDER, EU), the project H2020-MSCA-RISE-2017 Grant FunFiCO-777740, the project SEJI/2017/042 (Generalitat Valenciana), the Consolider Program CPANPHY-1205388, and the Severo Ochoa grant SEV-2014-0398 (Spain). This article is based upon work from COST Action CA15117, supported by COST (European Cooperation in Science and Technology).
%%%%%%%%%%%%%%%%%%%%%%%%%%%%%%%%%%%%%%%%%%%%%%%%%%%%%%%%%%%%%%%%%%%%%%%%%%%%%%%%%%%%

%%%%%%%%%%%%%%%%%%%%%%%%%%%%%%%%%%%%%%%%%%%%%%%%%%%%%%%%%%%%%%%%%%%%%%%%%%%%%%%%%%%%

%%%%%%%%%%%%%%%%%%%%%%%%%%%%%%%%%%%%%%%%%%%%%%%%%%%%%%%%%%%%%%%%%%%%%%%%%%%%%%%%%%%%

\end{document}